\documentclass[10pt]{article}
\usepackage{graphicx}

\def\Title#1{\begin{center} {\Large #1 } \end{center}}
\def\Author#1{\begin{center}{ \sc #1} \end{center}}
\def\Address#1{\begin{center}{ \it #1} \end{center}}

\newcommand\pubblock{\rightline{\begin{tabular}{l} Proceedings of the Second Annual LHCP\\ \pubnumber\\
         \pubdate  \end{tabular}}}

\newenvironment{Abstract}{\begin{quotation} \begin{center} 
             \large ABSTRACT \end{center}\bigskip 
      \begin{center}\begin{large}}{\end{large}\end{center} \end{quotation}}

\newenvironment{Presented}{\begin{quotation} \begin{center} 
             PRESENTED AT\end{center}\bigskip 
      \begin{center}\begin{large}}{\end{large}\end{center} \end{quotation}}

\def\beq{\begin{equation}}
\def\eeq#1{\label{#1}\end{equation}}
\def\eeqn{\end{equation}}

\def\beqa{\begin{eqnarray}}
\def\eeqa#1{\label{#1}\end{eqnarray}}
\def\eeqan{\end{eqnarray}}

\let\bar=\overbar

\def\Dslash{\not{\hbox{\kern-4pt $D$}}}
\def\dslash{\not{\hbox{\kern-2pt $\del$}}}

\def\msb{{\bar{\ssstyle M \kern -1pt S}}}

\usepackage{color}

\newcommand\pubnumber{ }

\newcommand\pubdate{\today}

\def\affiliation{
On behalf of the CMS collaboration, \\
Northeastern University, Boston, 02115, U.S.A }

\begin{document}

\large
\begin{titlepage}
\pubblock

\vfill
\Title{  VBS/VBF from CMS  }
\vfill

\Author{ Andrea Massironi }
\Address{\affiliation}
\vfill
\begin{Abstract}

Vector Boson Scattering (VBS) and Vector Boson Fusion (VBF) studies in pp collisions at 7 and 8 TeV center of mass energy
based on data recorded by the CMS detector at the LHC in 2011 and 2012 are reported.

\end{Abstract}
\vfill

\begin{Presented}
The Second Annual Conference\\
 on Large Hadron Collider Physics \\
Columbia University, New York, U.S.A \\ 
June 2-7, 2014
\end{Presented}
\vfill
\end{titlepage}
\def\thefootnote{\fnsymbol{footnote}}
\setcounter{footnote}{0}
%

\normalsize 


\section{Introduction}

Although the Standard Model of Particle Physics has been extensively tested at the
colliders, proving to be accurate over many order of magnitude in the prediction of processes
cross-sections, still some fundamental pieces need to be measured.
The question about the existence of the Higgs boson, has now been split into two queries:
is this the only Higgs boson present in Nature, and is it really doing the Higgs boson job.
To answer these questions, there are three main roads:
keep searching for scalars outside of m $=$ 125 GeV~\cite{Chatrchyan:2013yoa};
measure with high precision the properties of the 125 GeV boson~\cite{CMS:2014ega};
measure the electroweak vector boson interactions,
establishing whether this Higgs boson really can make unitarity-preserving the Vector Boson Scattering (VBS) amplitudes at all energies.
The test electroweak production of single Z boson, 
that is one of the first Vector Boson Fusion (VBF) processes measured at LHC~\cite{Chatrchyan:2013jya,CMS:2013qfa},
and the projections for VBS searches at 13 TeV~\cite{CMS-PAS-FTR-13-006}
performed by CMS experiment
are the topics of this report.

\section{Electroweak production of Z boson in association with two forward/backward jets}

\begin{figure}[h!tb]
\centering
\includegraphics[height=2.0in]{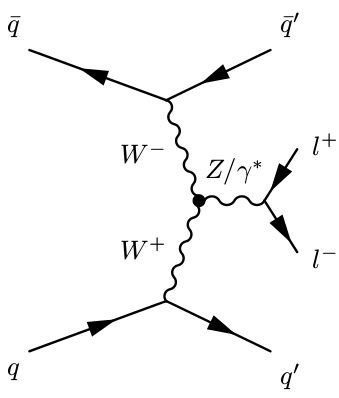}\ \ \ 
\includegraphics[height=2.0in]{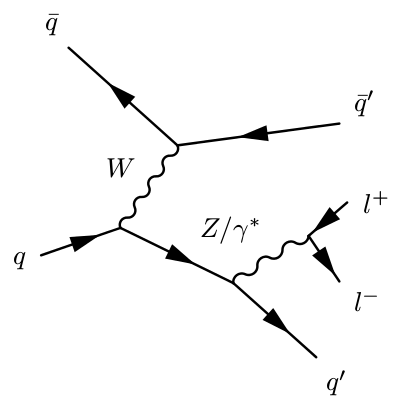}\ \ \ 
\includegraphics[height=2.0in]{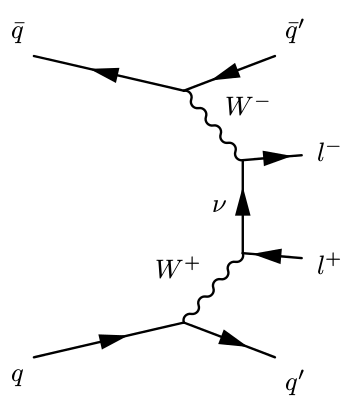}
\caption{
  Representative diagrams for dilepton production in association with two jets from pure electroweak processes.
  Vector boson fusion (left), Bremmstrahlung-like (center), and multiperipheral (right) productions.
  }
\label{fig:processes}
\end{figure}

\begin{figure}[h!tb]
\centering
\includegraphics[height=2.5in]{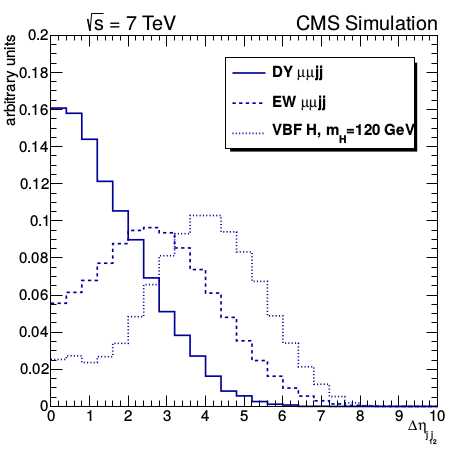}
\includegraphics[height=2.5in]{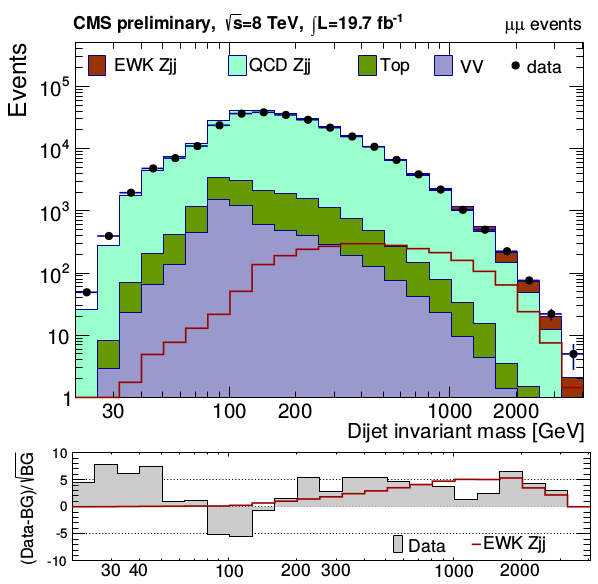}
\caption{
  Pseudorapidity separation between tag jets, $\Delta\eta_{jj}$ (left).
  The invariant mass of the two tag jets, $m_{jj}$ (right).
  }
\label{fig:detajjmjj}
\end{figure}

In proton proton collisions the dominant source of
production of a Z boson followed by a leptonic decay $Z \rightarrow \ell\ell$
in association with two jets
is through mixed electroweak (EW) and strong (QCD) processes of order
O($\alpha^2_{EWK}$ $\alpha^2_{QCD}$), commonly known as Drell-Yan (DY) plus two jets processes.
Pure electroweak productions of the $\ell\ell$jj
final state, of order O($\alpha^4_{EWK}$), are more rare at the LHC~\cite{Oleari:2003tc},
but are expected to carry a distinctive signature which can be explored experimentally:
two jets of very high energy, well separated
in pseudorapidity and with a large invariant mass, are expected to be produced in association
with the dilepton pair.
In the following, these jets that originate from the
fragmentation of the outgoing quarks in EW processes will be referred as ``tag jets'' and to the the process
which originated them as EW-Z+2jets.
Figure~\ref{fig:processes} shows representative Feynman diagrams for the
EW production of dilepton pairs in association with two jets including Vector Boson Fusion
(VBF) processes, Z-bremsstrahlung processes, and multiperipheral processes;
which have a large negative interference between each other.
However the VBF electroweak production can be isolated from other processes,
by means of selections based on the VBF topology,
that reduce also important backgrounds to the search.
One important background of this search is the QCD production of Z+2 jets.
As shown in Figure~\ref{fig:detajjmjj} (left), 
the pseudorapidity separation between the tag jets ($\Delta\eta_{jj}$)
is greater for electroweak production of Z+2 jets with respect to the QCD production mechanism.
Another important variable to distinguish between VBF Z production and other backgrounds is
the invariant mass of the two tag jets, $m_{jj}$, as shown in 
Figure~\ref{fig:detajjmjj} (right).

\section{Analysis strategy and results}

\begin{figure}[h!tb]
\centering
\includegraphics[height=2.7in]{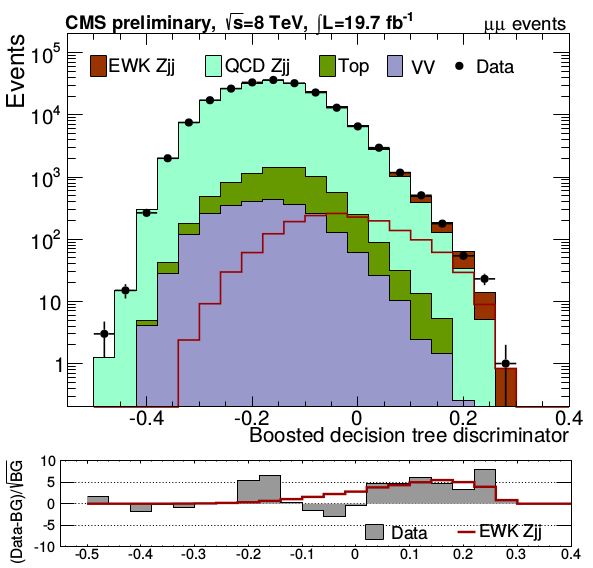}
\includegraphics[height=2.7in]{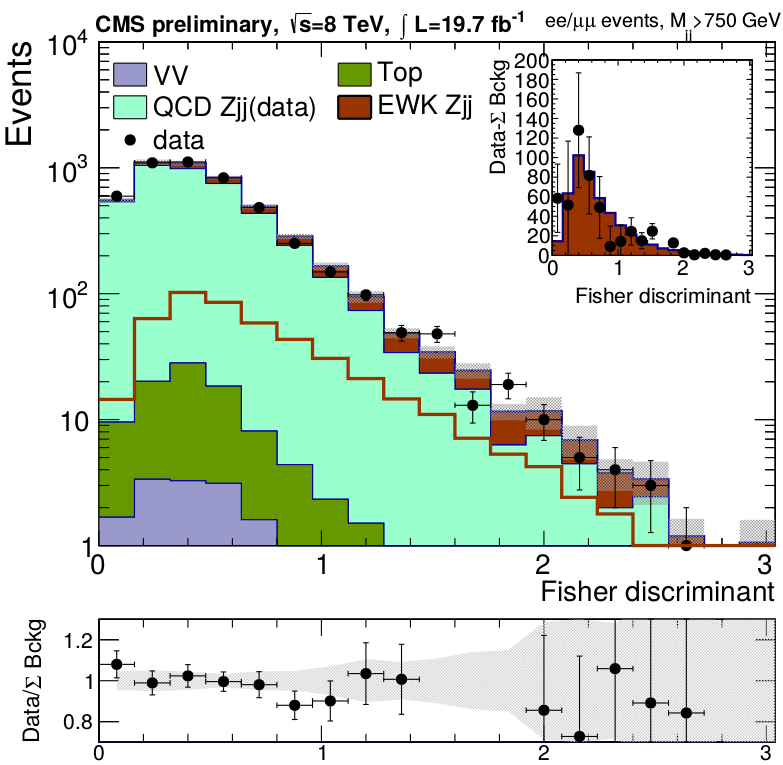}
\caption{
  The BDT output (left) after selecting the Z candidate and the tag jets.
  The expected contributions from the signal and background processes are evaluated from simulation.
  The right figure shows the linear discriminator shape in the most sensitive region ($m_{jj}$ $>$ 750 GeV)
  and the inset displays the result of a bin-by-bin background subtraction.
  The total uncertainty from the data-driven background prediction obtained from the photon control sample is shown
  as a shaded band in both the distribution and the bottom panel.
  }
\label{fig:bdtlin}
\end{figure}

Events with at least two charged leptons (electrons or muons) are selected in data using
dilepton triggers. Each lepton is required to have a transverse momentum $p_T>$~20~GeV and to be
reconstructed within $|\eta|<$~2.4, where $\eta$ is its pseudo-rapidity. Isolation criteria is furthermore
imposed by using the tracks which are reconstructed close to the leptons.
The sum of the transverse momenta of the tracks is required not to exceed more than 10\% of the
$p_T$ of each lepton candidate. In each event the leptons are required to have opposite charge and the
invariant mass of the system must be compatible with a Z boson, i.e. $|M_{ll}-M_Z|<$~15~GeV.
The two tag jets are selected within $|\eta|<$~4.7 and required to have $p_T>$~30~GeV.\\
Since the main background is QCD production of Z+2 jets, two different approaches
to have an estimation of its contribution are performed.
The first one is based on the use of MC expectation, 
propagating all theoretical and experimental uncertainties.
In this method a multrivariate discriminator based on jets $p_T$,
azymuthal and pseudorapidity distance between jets, 
invariant mass of the two tag jets, 
the azymuthal distance between jets and the Z boson, 
and the rapidity of the Z boson in the lab and in the 
di-jet frame,
is used.
Figure~\ref{fig:bdtlin} (left) 
shows the multivariate discriminator distribution, 
where the electroweak Z+2 jets contribution is clearly visible
in the high score region.\\
The second methos is based on the use of 
$\gamma$+jets sample from data to emulate
Z+jets after re-weighting the $p_T$
distribution between the photon and the Z boson.
The analysis is then performed,
in several $m_{jj}$ bins,
by means of a linear discriminant based on
$\Delta\eta_{jj}$, $m_{jj}$ and 
the ratio between the transverse momentum of the  di-jet system and the scalar sum 
of the transverse momentum of the two jets.
Figure~\ref{fig:bdtlin} (right)
shows the distribution of the linear discriminant:
a 5.9$\sigma$ (5.0$\sigma$) observed (expected) discovery significance has been established.

\section{Characterization of VBF topology}

\begin{figure}[h!tb]
\centering
\includegraphics[height=2.7in]{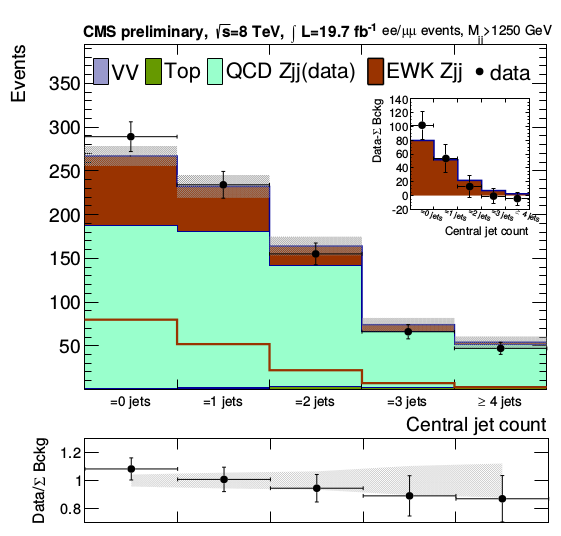}
\includegraphics[height=2.7in]{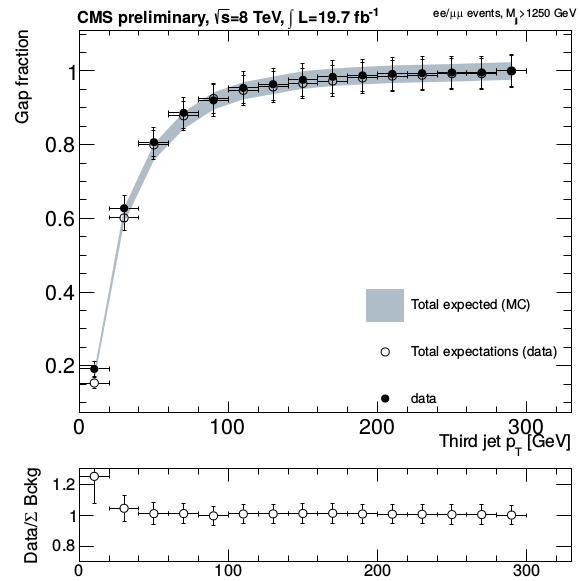}
\caption{
  Distribution of the number of reconstructed jets (left)
  and the jet veto efficiency as a function of the jet transverse momentum (right).
  The reconstructed quantities are compared directly to the prediction obtained with a full simulation of the CMS
  detector.
  }
\label{fig:distr}
\end{figure}

After establishing the signal, the properties of the hadronic activity in the selected events are studied: 
radiation patterns in Z+multijet events,  the charged hadronic activity as function of several kinematics variables,
and the production of extra jets in a high purity region ($m_{jj}$ $>$ 1250 GeV) have been measured.
Overall, a significant suppression of the hadronic activity for the signal is expected,
due to the fact that the final state objects have their origin in pure electroweak interactions
in contrast with the production of Z+2 jets via QCD. 
In Figure~\ref{fig:distr}, the distribution of the number of jets and jet veto efficiency are reported.
A good agreement between MC expectations and data is found.
Such variables and a deep understanding of the VBF topology
are crucial also as input for VBF Higgs searches,
where the correct prediction 
of signal and background distributions is very important.

\section{Future studies}

\begin{figure}[h!tb]
\centering
\includegraphics[height=2.5in]{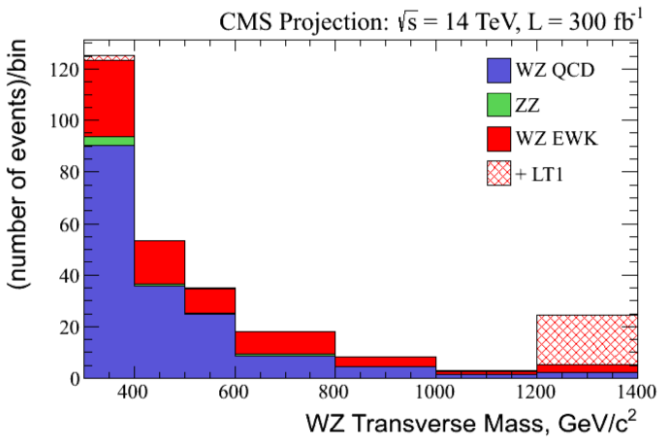}
\caption{
  The WZ transverse mass at 14 TeV with 300 fb$^{-1}$.
  The additional contribution of an anomalous couplings signal
  for the L$_{T1}$ operator (LT1) with f$_{T1}$ / $\Lambda^4$ $=$ 1.0 TeV$^{-4}$ is illustrated by the red hatched histogram.
  }
\label{fig:future}
\end{figure}

The LHC performances during run I are very successful,
and fruitful results have been accomplished.
In the preparation for LHC run II,
not being limited any more by statistics
thanks to higher cross section of rare processes
and higher instantaneous luminosity,
new vector boson scattering
processes, such as WZ electroweak production~\cite{CMS-PAS-FTR-13-006},
have been looked for.
An integrated luminosity of about 75 fb$^{-1}$ is sufficient for 3$\sigma$ evidence
of the WZ electroweak production,
while 185 fb$^{-1}$ is sufficient for 5$\sigma$ observation
at 14 TeV center of mass energy.
With 13 TeV center of mass energy, approximately 15\% more data
is needed for the discovery of the electroweak scattering process.\\
To search for new physics processes that result in 
anomalous couplings, the WZ transverse mass has been used for
final discrimination as the center of mass of the scattering system
is highly sensitive to new physics involving massive particles. 
For 300 fb$^{-1}$ the expected sensitivity to anomalous couplings is f$_{T1}$ / $\Lambda^4$ $=$ 0.8 TeV$^{-4}$.
Figure~\ref{fig:future} shows the transverse mass of the WZ system 
for the SM case and in case of anomalous couplings.

%

%




\section{Conclusions}

The LHC run I has been very successful.
The first vector boson fusions processes have been observed observed:
the electroweak Z + 2 jets production has been established with 
more than 5$\sigma$ significance.
So far, consistency between theory and data has been found.
A measurement of the properties of the hadronic activity in Z + 2 jets events is performed.
It is clear that although we discovered a Higgs boson,
the comprehension of the Electroweak Symmetry Breaking is not
completed, as several EW boson self-couplings still need to be observed.
To exhaustively complete the picture, 100 to 300 fb$^{-1}$ of integrated luminosity of data 
collected at $\sqrt{s}$ $=$ 13-14 TeV are needed.
First projections on the discovery potential of anomalous
gauge couplings have been performed.
The vector boson scattering and vector boson fusion processes
are going to be two of the hot topics at Run II of LHC.



\begin{thebibliography}{99}
\bibitem{CMS:2014ega} 
  CMS Collaboration, 
  ``Precise determination of the mass of the Higgs boson
and studies of the compatibility of its couplings with the standard model,''
  CMS-PAS-HIG-14-009.
\end{thebibliography}
\end{document}